\documentclass{appolb}
\usepackage{epsfig,amssymb,xspace,color}

\newcommand{\nc}{\newcommand}
\nc{\req}[1]{Eq.\,(\ref{#1})}  \nc{\eps}{\varepsilon}
\nc{\beq}{\begin{equation}}     \nc{\beql}[1]{\begin{equation}\label{#1}}
\nc{\eeq}{\end{equation}}       \nc{\rf}[1]{figure  \ref{#1}}
\nc{\beqa}{\begin{eqnarray}}   \nc{\eeqa}{\end{eqnarray}}
     \nc{\pathlaptop}{/home/rafelski/figure/}
     \nc{\pathletes}{/users/lpthe/jletes/bookraf/figures/}
     \nc{\pathnow}{}
 \def\lessim{\lower.5ex\hbox{$\; \buildrel < \over \sim \;$}}
\def\gtrsim{\lower.5ex\hbox{$\; \buildrel > \over \sim \;$}}

\usepackage{graphicx}

\begin{document}
\title{My strange times with Johann Rafelski
\thanks{Jubilee session, SQM2011, Krakow}}%
\author{Giorgio Torrieri
\address{FIAS,
  J.W. Goethe Universit\"at, Frankfurt am Main, Germany}}
\maketitle
\begin{abstract}
I will give a short review the physics of strangeness enhancement in quark-gluon plasma, and argue that it is currently the best candidate of a signature of deconfinement.   I will also discuss what strangeness abundance can tell us about the bulk properties of the system created in heavy ion collisions.  
\end{abstract}
\PACS{25.75.-q, 25.75.Dw, 25.75.Nq}
  
\section{Why is Jan here?}
I do not need to tell you why Jan's jubilee is celebrated at this conference.
We are here to discuss how strangeness, and more generally, hadronic flavor chemistry, can be used to study the thermodynamic properties of the system created in heavy ion collisions, and hopefully to determine the onset of quark deconfinement.   Jan was so instrumental in all of these topics that, most likely, without him these workshops would not exist, at least in their current form.

What does a hot strongly interacting system look like?   Before quantum chromodynamics was discovered, people thought hadrons were simply made of other hadrons, ``by their bootstraps''.   There was thought to be an infinite tower of hadronic states decaying and interacting with other states, and distinguishing ``fundamental'' from ``composite'' states would be impossible.
It follows that distinguishing a ``highly excited resonance'' from a ``fireball of hadrons'', a large system described in the previous paragraph, becomes impossible.  In a remarkable achievement, Rolf Hagedorn \cite{jansbook} showed that this {\em must imply} an ``ultimate temperature'' beyond which this description breaks down.  This is where Jan Rafelski first entered this field, demonstrating, in collaboration with Hagedorn, that this temperature coincides with the temperature at which quarks deconfine into hadrons.   Hadro-chemistry therefore becomes a probe into the thermal conditions of the system, capable of showing how close does the hadronic freeze-out happen wrt the Hagedorn temperature \cite{hagedorn}.

To go further, other simplifications are necessary.   For instance, the light quark's mass is so small with respect to the critical temperature $T_c$ that, in a plasma of quarks and gluons, it should behave essentially as a massless fermion.   At the other end of the spectrum, the heavy charm and bottom quark mass is so large wrt $T_c$ that $T/m_{c,b}$ can be thought of as a small parameter to expand around.
Intriguingly,  the strange quark is exactly inbetween.   Its mass is of the order of $T_c$ so, in a quark gluon plasma just above deconfinement, the strange quark is neither light nor heavy.
Jan Rafelski's insight is that this ``bug'' is actually a feature, allowing us to use strange quarks {\em to clock} the evolution of the system when it is in a deconfined state.   

Let us think how strange quarks would behave in a thermalized plasma of light quarks and gluons, vs how strange hadrons would behave in a thermalized gas of hadrons at a similar temperature $\sim T_c$:  First of all, making $s \overline{s}$ pairs will be much easier in a hadron gas through $gg \leftrightarrow s \overline{s}$ collisions ($\sqrt{s} \geq 2 m_s \sim T_c$) then through hadronic reactions such as $n \pi \leftrightarrow \Lambda K$ ($\sqrt{s} \sim m_K + m_\Lambda-m_\pi-m_N \sim 500 MeV \gg T_c$).    Hence, the timescale for chemical equilibration of strange quarks will be much faster than the timescale for equilibration of strange hadrons.
Because $(m_s-m_q)/T_c \gg (m_K-m_\pi)/T_c$, the {\em equilibrium abundance} of such strange quarks will also be greater in a Quark-Gluon plasma \cite{orig1,orig2}.

These arguments lead to the realization that, for a nuclear event where a QGP was formed, the strangeness abundance will be considerably greater than for a ``similar'' event with no QGP.    If hadronization happens ``quickly'' with quark recombination, this extra abundance will manifest itself with a large enhancement of {\em multi-}strange hadrons such as the $\Omega$, since these can only be produced by sequential interactions such as $N \pi \pi \pi \leftrightarrow \Lambda K \pi \pi \leftrightarrow \Xi K K \pi \leftrightarrow \Omega K K K$ in a hadron gas.
Hence, experimentally measuring enhancement is a diagnostic of QGP formation in heavy ion collisions.  The enhancement of $Y$ can be experimentally defined as $N_{Y}^{AA}/(R N_{Y}^{pp,pA})$ where $R$ is a normalization which is either $N_{part}^AA/N_{part}^{pp,pA}$ or $(dN/dy)_{AA}/(dN/dy)_{pp,pA}$
\section{Why Jan should be happy}
\begin{figure}[h]
\begin{center}
\epsfig{width=12cm,figure=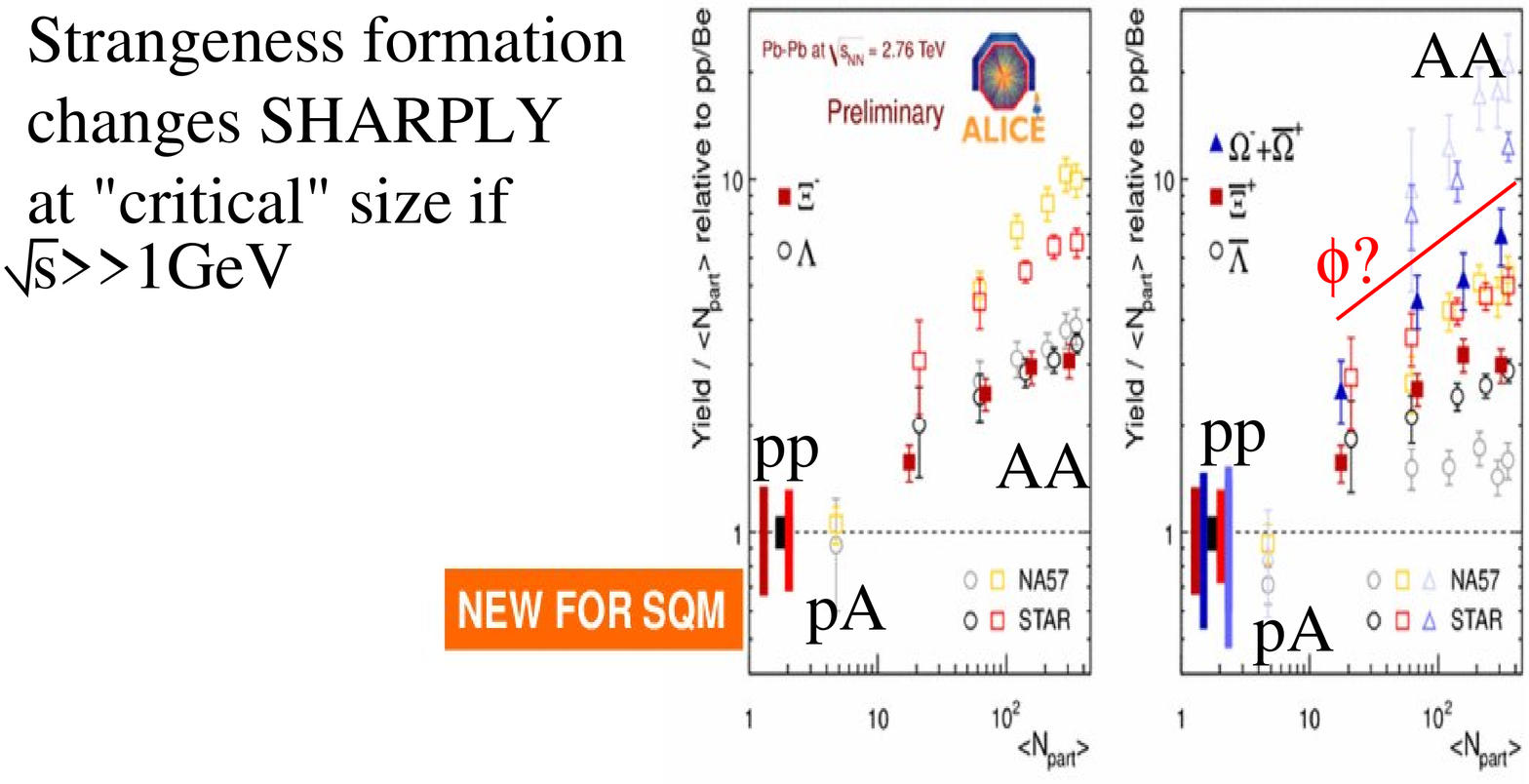}
\caption{\label{enhancement} Strangeness enhancement at the LHC and lower energies}
\end{center}
\end{figure}
This is a situation where the term ``a picture is worth $10^n$ words'' should apply.  The picture in question is Fig. \ref{enhancement} \cite{alice}.   
As can be seen, at all energies much higher than the Coulomb barrier energy, $A-A$ strangeness abundance is considerably enhanced wrt either $p-p$ or $p-A$.
The crucial question is whether this enhancement is due to ``chemistry'' (more strange particles per unit volume) or ``kinematics'' (the necessity to conserve strangeness exactly, which suppresses strange quark abundance in smaller systems) \cite{tounsi}.  This question can be answered by including the $\phi$ on this plot, something generally not done but which  {\em should} be, as \cite{sqm07} it is fundamental to clarify the physical origin of enhancement.
$\phi$s are strangeness-neutral, and hence are immune from any additional suppression due to ``canonical effects'' (the necessity to exactly conserve strangeness, and the difficulty to do so in smaller systems).
Thus, if strangeness enhancement is actually due to canonical suppression in $p-p$ collisions, one would expect no enhancement of the $\phi$, as well as a plateau once a ``large system size'', where canonical corrections become unimportant.
Exactly the opposite behavior is observed at {\em all} energies in Fig. \ref{enhancement}, showing the bulk of the enhancement is due to a different chemical content rather than a change in conservation law constraints.
Looking at the difference between $p-p,p-A$ and $A-A$ in Fig. \ref{enhancement}, it is difficult not link this change to a phase transition, so sudden it is.
\begin{figure}[t]
\begin{center}
\epsfig{width=14cm,figure=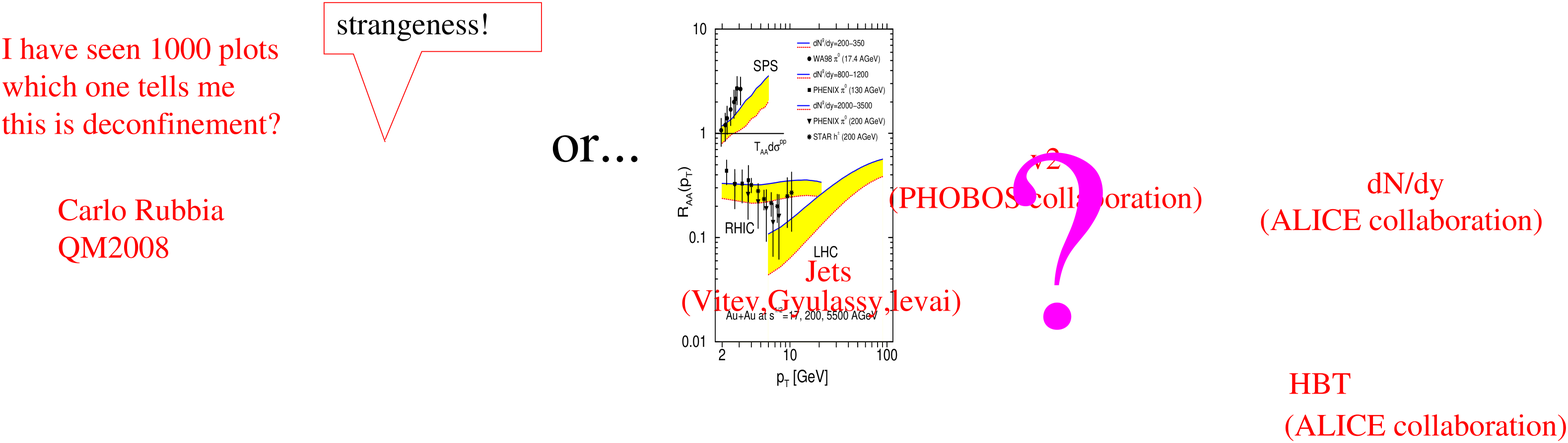}
\caption{\label{rubbia} A notable moment in QM2008 (left panel), and Jan's reply (center panel) had he been present at that conference.  The right-hand panels shows a sampling of heavy ion experimental results, demonstrating no obvious sign of a violation of scaling indicating a change in degrees of freedom}
\end{center}
\end{figure}
So, is strangeness enhancement really a deconfinement signature?  While it {\em looks like it}, the evidence is so far {\em not} conclusive.
The definite signature would be for the effect seen in Fig. \ref{enhancement} to {\em turn off}, for enhancement to disappear in A-A collisions at energies where any effect of the QGP phase is nonexistent or negligible.   This has not been seen, but {\em might be} in future lower energy experiments \cite{fair,shine,rhiclow,nica}.
Kinematically, the $\Omega$ mass is well below the scale at which the Coulomb barrier becomes important, so there is plenty of opportunity for this exploration.
Experiments such as \cite{fair,shine,rhiclow,nica} will be capable of measuring rare probes such as $\phi,\Xi,\Omega$ particles to high precision.

While how strangeness enhancement turns on is still an open question, it is worth to stop and contemplate that {\em other} observables do not yield a scaling violation which is nearly as clear as in Fig. \ref{enhancement} (Fig. \ref{rubbia} right panel).  In fact, their scaling with energy and system size is remarkably smooth, with no obvious hints of transition, immediately apparent in strangeness enhancement graphs.   This of course does not mean other observables are unworthy of study, rather the opposite (the scalings in the right panel of Fig. \ref{rubbia} are interesting, profound, and largely unexplained).   It is however worth remembering that physicists in other fields do have a habit of asking us which, of the hundreds of elaborate graphs our field has produced, tells us that ``this is the QGP'' (Carlo Rubbia asked this question in a plenary session of QM2008, as seen in the left and center panel of Fig \ref{rubbia}).  The answer to this question is still not conclusively there, but strangeness enhancement is certainly the best candidate.  Jan, therefore, has every reason to be happy.
\section{Why Jan is not always happy}
Given this spectacular success, one would expect Jan to smugly sit on his laurels, rather than raise hell at every experimental talk and competing theoretical talk he encounters.   

 In this section, I will attempt to show that this behavior actually also has a good scientific explanation.
For Jan, the issue has never just been the {\em existence} of strangeness enhancement.  He wants to use strangeness enhancement as a {\em tool} to characterize the bulk properties of the system created in heavy ion collisions.

The simplest way to do this is  to incorporate strange particles into a thermal analysis fit, to try to extract the temperature and baryochemical potential $\mu_B$.
  The goodness of your fit would confirm that, as Jan predicted in \cite{orig1}, the strangeness suppression factor , $\gamma_s \simeq 1$ so strangeness is a part of the equilibrated properties of the system. Therefore, talking about ``strange'' vs ``non-strange'' thermal properties is redundant. {\em all} particles will go into the partition function to compare data to the equation of state \cite{becattini,kaneta,pbm}.

This is the {\em simplest} approach, but it is not necessarily the {\em physically correct } one, and might make you forget something crucial.
If strangeness and entropy were equilibrated in a QGP, it is not at all certain they will also equilibrate in the HG {\em hadronizing} from the QGP.  They of course will if chemical equilibrium is maintained around $T_c$, but this is far from guaranteed, especially since, even in a cross-over regime, the ``width of $T_c$'' is so narrow that an expanding system will cross it in less than a $fm$ \cite{jansbook}.
 The degree of chemical equilibrium in the system can be ascertained by whether the dimensionless variable 
\begin{equation}
\alpha = \left[\frac{\chi_{s}(T,\mu_B)}{\rho_s(T,\mu_B)}\right]\left[\frac{d T}{d \tau} + \frac{d \mu_B}{d \tau}\right] \tau_{s}
\label{alphaeq}
\end{equation}
is $\ll 1$ (equilibrium is maintained) and $\geq 1$ (equilibrium breaks down).
$\chi_s$ and $\rho_s$ can be measured on the lattice.   The rate of change of $T$ and $\mu$ can be read from hydrodynamics.  $\tau_s$, the chemical equilibration timescale, is however unknown around $T_c$, and could diverge if the bulk viscosity (tracking the timescale of chemical equilibration) diverges \cite{mueller}.

If it diverges, then one {\em can not} use jut $T$ and $\mu$ because the hadron abundances will reflect the quark abundances of {\em the QGP} system, {\em not} of the equivalent equilibrium hadron system.
One has to additionally use the parameters $\gamma_{s,q}$, denoting lack of equilibrium of strange and light quarks.
If this is correct, you would expect these parameters to be $>1$ (an {\em impossible} result to obtain from a quasi-particle transport model, unless put in as an initial condition), because both entropy and strangeness are higher in a QGP than in a HG.
And \cite{petran}, this is exactly what the data seems to say if $\gamma_{q,s}$ is included in fits.
\begin{figure}[h]
\begin{center}
\epsfig{width=14cm,figure=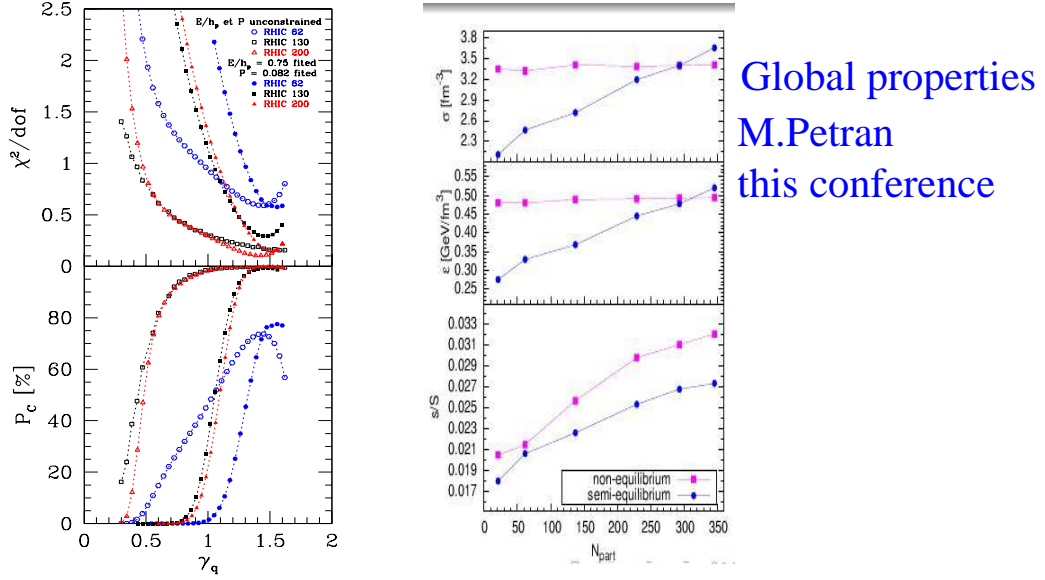}
\caption{\label{chiprof} Left panel: $\chi^2$ profiles for $\gamma_q$ Right panel: Ratios of bulk observables in the equilibrium and non-equilibrium scenario \cite{petran}}
\end{center}
\end{figure}
Currently,  as Fig. \ref{chiprof} (left panel) shows, fitting can not tell you whether $\gamma_{q,s}$ are really physically necessary parameters or fudge factors.   The statistical significance difference between equilibrium and non-equilibrium is simply not enough.

If you believe statistical models necessarily imply equilibration, you will automatically reject $\gamma_{q,s}$ as unphysical.  If you buy Jan's arguments, you will accept $\gamma_{q,s}$ after the best fit value also fits the expected ratio of bulk abundances at hadronization (right panel of Fig. \ref{chiprof}. The corresponding equilibrium curves vary strongly in energy and system size, suggesting freeze-out densities are not related to any fundamental QCD value )
\section{And what we can do about it}
Well, the simplest thing is not to assume the simplest model is automatically the right one.   The sudden hadronization chemical non-equilibrium scenario {\em could well} be incorrect, but it is {\em has} physical justification.  It therefore needs to be considered as a {\em distinct theory} from the equilibrium scenario, even through it has more parameters.   The question is whether these parameters are ``real'' or simply fudge factors.   First of all, one needs a framework in which all statistical models can be analyzed in an ``objective'' way.  Constructing such a framework \cite{share,sharev2} was the focus of my thesis, and of my long collaboration with Johann as well as the organizers of this conference.

Then, ideally, one should be able to construct observables more sensitive to $\gamma_q$.
Fluctuations immediately come to mind, because higher $T$ tends to lower them, while higher $\gamma$ tends to raise them due to Bose-Einstein corrections.  Hence, unlike for yields, $T,\gamma$ are {\em anti} correlated and a wrong value for fluctuations will fail at describing them simultaneously \cite{sharev2}. 
A systematic comparison to experimental data for this is hopefully right around the corner.
\section{Conclusions}
The appropriate conclusion is that we all know why Jan is here.
We still have to find out if Jan is right about everything or not, but
strangeness enhancement has for sure made an enormous impact on the field, and remains the premier ``smoking gun'' candidate for deconfinement.

The only question worth a mention is why should I, of all people, give this talk about it?   Basically, in  1997 I started my PhD on the NA57 experiment, to measure strangeness abundance in p-p,p-A and A-A collisions After three years, I realize I am just not made for experiment, but I really loved this field.   
So I quit with a masters, and start writing to various theorists whose work I learned during the course of my studies.  Jan Rafelski immediately comes to mind.   He is also the one who answers the email.  This talk describes some of what happened next.  I am here because I am grateful for the opportunity that Jan offered me. I hope his faith in me was at least a bit justified.

G.~T.~acknowledges the financial support received from the Helmholtz International Centre for FAIR within the framework of the LOEWE program
(Landesoffensive zur Entwicklung Wissenschaftlich-\"Okonomischer Exzellenz) launched by the State of Hesse, the organizers of SQM for their generous
support, and Johann Rafelski for the years of collaboration, discussion, and support given.

\end{document}